\documentclass[prb,twocolumn,amsmath,amssymb,superscriptaddress,floatfix,nofootinbib]{revtex4-2}
\usepackage{epsfig, graphicx,graphics,amsmath,amssymb,float}
\usepackage[T1]{fontenc}
\usepackage[latin9]{inputenc}
\usepackage{amsmath}
\usepackage{amssymb}
\usepackage{appendix}
\usepackage{amscd}
\usepackage{bm}
\usepackage{psfrag}
\usepackage{bbm} 
\usepackage{babel}
\usepackage{wasysym}
\usepackage{mathrsfs}
\usepackage{color}
\usepackage{hyperref}
\usepackage[normalem]{ulem}

\begin{document}

\title{Boundary-induced Majorana coupling in a planar topological Josephson junction}

\author{Hyeongseop Kim}
\affiliation{Department of Physics, Korea Advanced Institute of Science and Technology, 
Daejeon 34141, Korea}

\author{Sang-Jun Choi}
\email{Present address: Department of Physics Education, Kongju National University, Gongju 32588, Republic of Korea.}
\affiliation{Department of Physics, Korea Advanced Institute of Science and Technology, 
Daejeon 34141, Korea}

\author{H.-S. Sim}
\email{hs_sim@kaist.ac.kr}
\affiliation{Department of Physics, Korea Advanced Institute of Science and Technology, 
Daejeon 34141, Korea}

\author{Sunghun Park}
\email{sunghun.park@ibs.re.kr}
\affiliation{Center for Theoretical Physics of Complex Systems,
Institute for Basic Science (IBS), Daejeon 34126, Republic of Korea}

\begin{abstract}
Understanding environmental effects in a topological Josephson junction is vital for identifying
signatures of Majorana modes. We consider a planar Josephson junction formed on
the surface of a three-dimensional topological insulator, which possesses Majorana modes inside the junction
and boundary modes outside. We find that tunneling between the inner and outer modes gives rise to effective coupling between the inner Majorana modes, and hence induces energy splitting of their states even in the absence
of the direct spatial overlap of their wave functions. The energy splitting is obtained analytically in the weak tunneling limit and is numerically
investigated for an arbitrary tunneling strength. We discuss in detail the evolution of the energy splitting
with an external perpendicular magnetic field and its effect on the shape
of the Fraunhofer pattern.  
\end{abstract}
\maketitle

\section{Introduction}
A hybrid structure of a three-dimensional topological insulator (3D TI) and an s-wave superconductor (SC) can exhibit topological superconductivity at the TI-SC interface as a consequence of the spin-orbit coupling in the TI surface states and the superconducting proximity effect \cite{Hasan_and_Kane_RMP2010, Beenakker_AnnualReview2013, Aguado_NuovoCimento2017}. Theoretically, Majorana zero modes (MZMs) appears at the core of half-quantum vortices of the TI-SC interface or in a Josephson junction made up of two topological superconductors \cite{Fu_and_Kane_PRL2008}. 
Signatures of MZMs in such structures have been examined by a conductance peak at zero bias in tunneling spectroscopy \cite{Hao-HuaSun_and_Jin-FengJia_PRL2016} or a fractional ac Josephson effect induced by a $4\pi$-periodic current-phase relation in microwave spectroscopy \cite{Wiedenmann_NatCommun2016}.

Planar geometry Josephson junctions (JJs) have been intensively studied for the last decade, 
as they serve flexible structures in device design adequate for investigating MZMs. 
SC-TI-SC planar JJs have a controllable knob, the superconducting phase difference between two superconductors, enabling us to create and detect MZMs in the single JJ \cite{Wieder_PRB2013, Sochnikov_NanoLett2013, Kayyalha_npj_quantum_materials2020,choi2019josephson,Abboud_UIUC_PRB2023}, and their extensions to Josephson multijunctions \cite{Kurter_NatCommun2015, GuangYang_PRB2019} were proposed to realize a network of JJs in which one can perform operations of MZMs \cite{Okugawa_PRB2022}. 
Also, the study of SC-normal-SC JJs with spin-orbit coupling in the normal metal is \textcolor{blue}{a} rapidly growing research field, both theoretically \cite{Hell_PRL2017, Pientka_PRX2017} and experimentally \cite{Fornieri_Nature2019, H.Ren_Nature2019}, as this system can achieve a topological phase transition into 1D topological superconductivity with a lower magnetic field compared to a nanowire-based hybrid device. 

For the application to topological quantum computation, it is crucial to understand the stability of MZMs against external perturbations such as temperature or quantum fluctuation effects \cite{Meng_and_Lutchyn_PRB2015} and coupling to boundary modes. In particular, the latter effect of the coupling to boundary modes was disregarded previously, although it inevitably occurs in experiments, for example, with planar JJs \cite{Williams_PRL2012}, and it is still not clear whether the boundaries of such JJ devices affect their stability and experimental signatures.   

Here, we consider a planar JJ on the surface of a 3D TI, where two $s$-wave SCs are linked by the TI surface, and concentrate on the effect of gapless boundary modes residing along the boundaries of the SCs outside the junction. Differently from previous studies focusing only on MZMs inside the junction, 
we study the effect of the boundary modes on the energy spectrum and the Josephson effect of the junction, taking into account of tunnel coupling between the MZMs and the boundary modes. 
We find that the coupling can give rise to an indirect route to lift the degeneracy of the states formed by the MZMs. 
The MZMs are hybridized into a complex fermion with a finite energy. This energy splitting is 
independent of the length of the boundary $L_b$ in the weak coupling limit and for the moderate case it is much more slowly decaying with $L_b$ than the exponential decay, with $L_J$, of the overlap between MZMs within the junction $\propto e^{-L_J/\xi}$ where $L_J$ is the junction length. In the realistic situation, $L_b$ and $L_J$ have a similar order of magnitude, hence the effective coupling between the MZMs by the boundary modes can play an important role.  

We obtain the analytic form for the boundary-induced energy splitting in the weak tunneling limit. We also calculate numerically the energy spectrum as a function of the phase difference for an arbitrary tunneling strength, which shows a transition from $4\pi$- to $2\pi$-periodicity.
Moreover, we investigate how the energy spectrum evolves with an external magnetic field which generates the modulation of the phase difference along the junction. In particular, if a flux quantum is inserted into the junction, the modulation creates a pair of MZMs, one is movable within the junction by varying the phase difference and the other is delocalized along the boundary. The boundary mediated hybridization between the MZMs is analyzed when the movable MZM approaches to the end of the junction. Finally, we discuss the effect of the boundary on the shape of the Fraunhofer pattern.

The paper is organized as follows.  
We present in Sec. \ref{sec2} the main results by considering a minimal model for tunnel coupling between the boundary modes and the MZMs and obtain an analytical form for the boundary-induced energy splitting in the weak coupling limit. Section. \ref{sec3} provides the model Hamiltonian and the details of the calculation of the energy spectrum of the junction. In Sec. \ref{sec4}, we discuss the effect of the boundary on Fraunhofer pattern, and summarize in Sec. \ref{conc}. 

\section{Boundary-induced Majorana coupling}\label{sec2}
\begin{figure}
    \centering
    \includegraphics[width=0.5\textwidth]{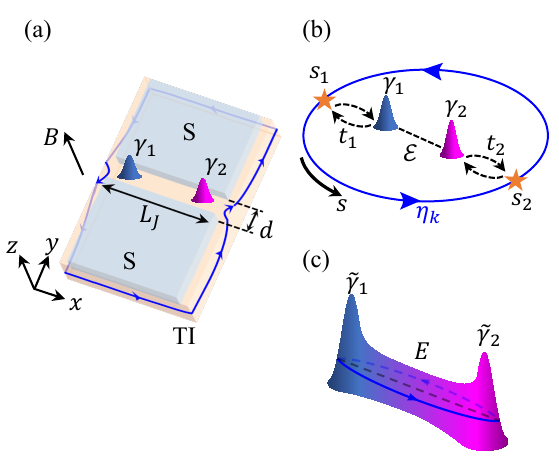}
    \caption{(a) A planar 3D TI JJ. Boundary channels (denoted in blue) affecting the junction are formed in the TI region which is not covered by the superconductors. Majorana modes can appear inside the junction or at the ends of the junction. (b) Schematics of Majorana modes $\gamma_1$ and $\gamma_2$ coupling with the boundary modes. (c) As the result of the coupling, the system is described by two new Majorana modes $\tilde{\gamma}_1$ and $\tilde{\gamma}_2$ with an effective coupling energy $E$. }
    \label{fig:SetUp}
\end{figure}
We present a Josephson junction device with the coupling between MZMs and boundary modes illustrated in Fig. \ref{fig:SetUp}(a). 
The device contains a topological JJ hosting two MZMs, $\gamma_1$ and $\gamma_2$, at the ends of the junction and gapless boundary modes outside the junction. The boundary modes are assumed to exist in the TI surface region, which is not covered by the superconductors, or along the edges of a finite-size 3D TI. Without loss of generality, we decompose the boundary modes into two chiral modes so that one of them, denoted by the loop drawn in blue in Fig. \ref{fig:SetUp}(a), couples with MZMs and the other remains intact. Figure. \ref{fig:SetUp}(b) shows a low-energy minimal model described by the Hamiltonian $H_{\text{min}} = H_M + H_B + H_T$, where 
$H_M=i\mathcal{E}\gamma_{1}\gamma_{2}$ is the Hamiltonian for the two MZMs with the coupling energy $\mathcal{E}$ due to their wave function overlap along the junction of length $L_J$, and $H_B = \hbar v_b k \,\eta_{k}^{\dag}\eta_{k}$ is for the single chiral mode of $\eta_k$ along the boundary with length $L_b$ and velocity $v_b$. Using the quantization condition of the loop $k L_b + \pi = 2\pi$ with the Berry phase $\pi$ originated from the spin rotation of the surface state along the round trip due to the spin-momentum locking, we find $k= \pi/L_b$. The tunnel coupling between the MZMs and the boundary mode occurring at positions $s_1$ and $s_2$ with the tunneling strengths $t_1$ and $t_2$, respectively, shown in Fig. \ref{fig:SetUp}(b) is given by 
\begin{equation}
H_T= \sqrt{2}t_{1}\gamma_1\eta_{k}+\sqrt{2}t_{2}\gamma_2\eta_{k}+\text{H.c.}    
\end{equation}

The key result of our study is that this coupling gives rise to an effective coupling between the inner MZMs. It can be shown by solving the Hamiltonian in the weak tunneling limit.   
It is then expressed by 
\begin{equation}
H_{\text{min}} = i E\,\tilde{\gamma}_1\tilde{\gamma}_2, 
\end{equation}
where $E$ is the effective coupling induced by the boundary mode illustrated in Fig. \ref{fig:SetUp}(c),  
\begin{equation}\label{effectiveE}
    E= \mathcal{E}+ \frac{|t_{1}+it_{2}|^{2}}{\mathcal{E}-\hbar v_b k}+ \frac{|t_{1}-it_{2}|^{2}}{\mathcal{E}+\hbar v_b k}+O(t^4),
\end{equation}
and the Majorana-boundary hybrid modes $\tilde{\gamma}_i$ are obtained by 
\begin{eqnarray}
    \tilde{\gamma}_1 =\gamma_{1}+\Gamma_{12} \eta_k + \Gamma^*_{12} \eta^{\dagger}_k\, , \\
    \tilde{\gamma}_2 =\gamma_{2}+\Gamma_{21} \eta_k + \Gamma^*_{21} \eta^{\dagger}_k\, ,
\end{eqnarray}
where 
\begin{equation}
    \Gamma_{mn} = \frac{\sqrt{2}\left(t_{m}\hbar v_b k+ (-1)^{n} it_{n}\mathcal{E}\right)}{\mathcal{E}^2-(\hbar v_b k)^2}.
\end{equation}
The result shows that the Majorana modes describing the system are superpositions of $\gamma_{1,2}$ and the boundary channels $\eta_{k}$ whose contribution increases as the coupling $t_{1,2}$ increase. The normalization factor of the boundary mode state $1/\sqrt{L_b}$ allows us to rewrite the coupling as $t_i = t'_{i}/\sqrt{L_b}$ where the parameter $t'_{i}$ is the length-independent coupling strength. This is because the coupling is proportional to the overlap between the Majorana state and the boundary mode state at the tunneling points. Then, for the case of $\mathcal{E}=0$, the effective coupling in Eq. \eqref{effectiveE} can be expressed as 
\begin{equation}
    E= \frac{4}{\hbar v_b \pi} \text{Im}\left[t'^{*}_{1} t'_2\right],
\end{equation}
which is independent of the length $L_b$. 

When one of the coupling is turned off, the decoupled Majorana mode is localized in the junction. For $t_{1}\rightarrow 0$, $\mathcal{E}\rightarrow 0$, then $E\rightarrow 0$ with $\tilde{\gamma}_1=\gamma_1$ and 
\begin{equation}
    \tilde{\gamma_{2}}=\gamma_2-\frac{\sqrt{2}t_{2}}{\hbar v_b k}\eta_{k}-\frac{\sqrt{2}t_{2}^{\ast}}{\hbar v k}\eta_{k}^{\dag}\,\, ,
 \end{equation}
 while for $t_{2}\rightarrow 0$, $\mathcal{E}\rightarrow 0$, then $E\rightarrow 0$, $\tilde{\gamma}_2=\gamma_{2}$, and
 \begin{equation}
     \tilde{\gamma_{1}}=\gamma_1-\frac{\sqrt{2}t_{1}}{\hbar v_b k}\eta_{k}-\frac{\sqrt{2}t_{1}^{\ast}}{\hbar v_b k}\eta_{k}^{\dag}\,\,.
 \end{equation} 
This situation happens in TI Josephson junction with magnetic field perpendicular to the junction
(see Sec. \ref{sec3.2} ).

\section{Theoretical model}\label{sec3}
\begin{figure}
    \centering
    \includegraphics[width=0.5\textwidth]{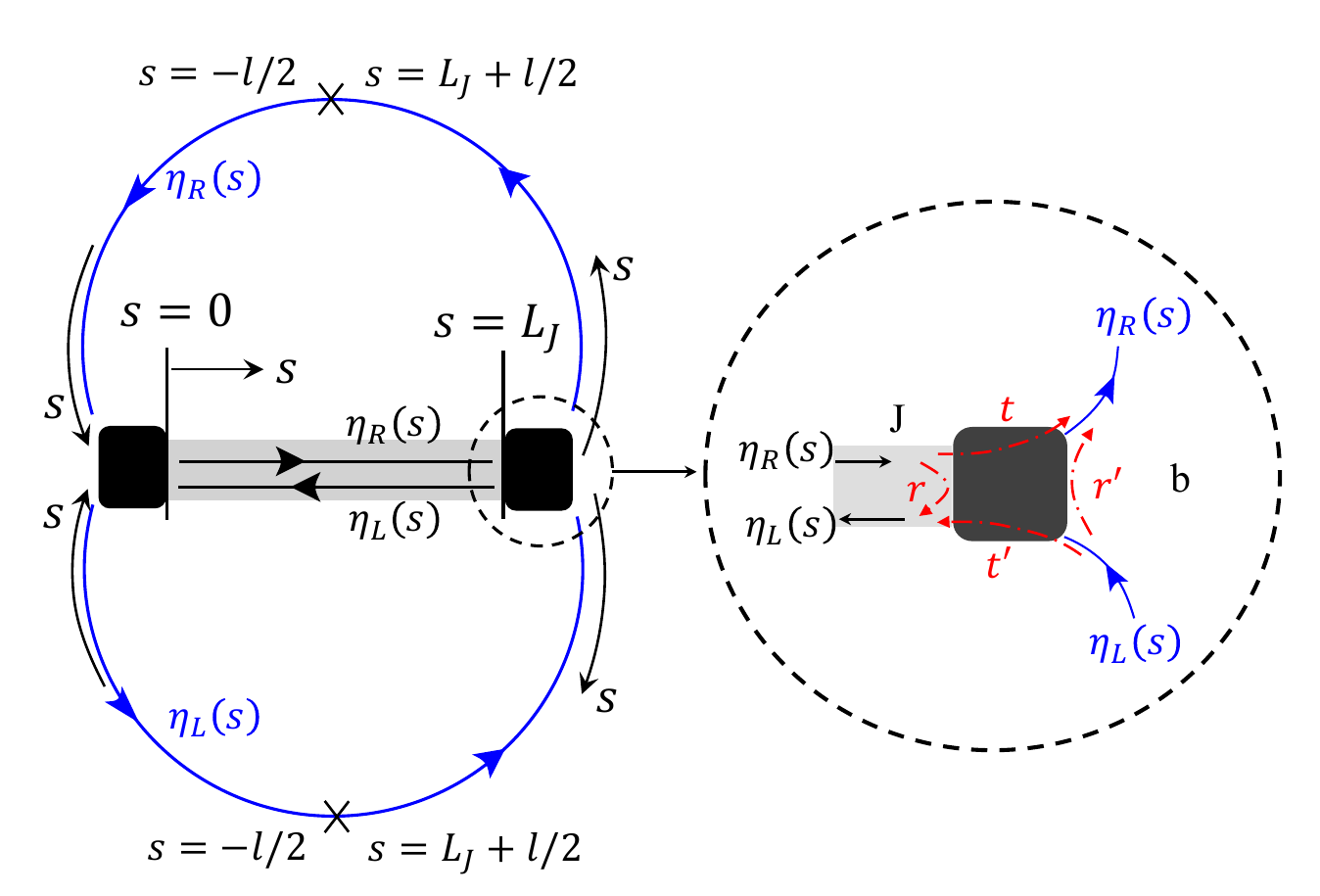}
    \caption{Left: An effective 1D model for the JJ (depicted by gray region) and its boundary channels (propagating along blue arcs arrows) showing in Fig. \ref{fig:SetUp}(a). 
    The arc coordinate $s$, which is aligned to the $x$-axis of the 2D Hamiltonian, Eq. \eqref{eq: 2d Hamiltonian}, is marked by dark arrows. 
    The direction of the arc coordinate $s$ is defined to be opposite to the propagation direction of the lower arc channel, hence the boundary channels of the upper and lower arcs are treated as the left and right moving channels of the 1D effective model in Eq. \eqref{eq: Effect 1D total H}.  
    Right dashed circle: The coupling region between the junction channels and the boundary channels is depicted by the dark box. The coupling is described by the scattering matrix elements $t, t', r, r'$ (red dashed-dot arrows). These elements are tuned by a parameter $U$ (see the text).
    $U\rightarrow\infty$ results in $|r'|^2=|r|^2=R=1$, which means that the boundary and junction channels are decoupled. 
    }
    \label{fig: 1d model}
\end{figure}
The planar JJ shown in Fig. \ref{fig:SetUp}(a) is made by proximity induced superconductivity on the surface states of 3D TI and is described by 2D Bogoliubov-de Gennes Hamiltonian $H=\frac{1}{2}\int d^2\mathbf{r}\Psi^{\dag}(\mathbf{r})\mathscr{H}(\mathbf{r})\Psi(\mathbf{r})$ with
\begin{equation}\label{eq: 2d Hamiltonian}
    \mathscr{H}(\mathbf{r})=\tau_{z}\left[ v_F \hat{z}\cdot\left(\vec{\pi}\times \vec{\sigma}\right)-\mu(\mathbf{r})\right]+\Delta(\mathbf{r})\tau_{+}+\Delta^{\ast}(\mathbf{r})\tau_{-}.
\end{equation}
Here $\Psi=(\psi_{\uparrow},\psi_{\downarrow},\psi_{\downarrow}^{\dag},-\psi_{\uparrow}^{\dag})^{T}$ is the Nambu spinor of the surface states and $\pi_{j}=-i\hbar\partial_{j}+eA_{j}(\mathbf{r})$ $(j=x,y)$ where the vector potential $\vec{A}(\mathbf{r})$ describes the magnetic field applied perpendicular to the surface by $\nabla\times \vec{A}=B(\mathbf{r})\hat{z}$. $\sigma_{j}$ and $\tau_{j}$ are Pauli matrices acting on the spin and particle-hole block respectively, and $\tau_{\pm}=(\tau_{x}\pm i\tau_{y})/2$.
The induced SC gap $\Delta(\mathbf{r})$ is $|\Delta (\mathbf{r})|=\Delta_0$ in the region $0<x<L_J$ and $|y|>d/2$, and zero elsewhere, where $L_J$ and $d$ are the width and the length of the JJ, respectively.  
The magnetic field in the junction region is set to be $B(y)=B_0$ and decays exponentially into the superconducting region with the distance from the superconducting interface characterized by the London penetration depth $\lambda_{L}$.
We note that the Zeeman energy is neglected in our model, as we are interested in the regime of sufficiently weak magnetic fields. The Zeeman energy only slightly renormalizes the parameters of our model (such as the Josephson coupling $m$ and the velocity of the boundary mode) in Eq. (\ref{eq: Effect 1D total H}) below, when the chemical potential is much larger than the Zeeman energy \cite{Fu_and_Kane_PRB2009, FuKane_PRL2009_Probing}.

\begin{figure*}
    \centering
    \includegraphics[width=0.95\textwidth]{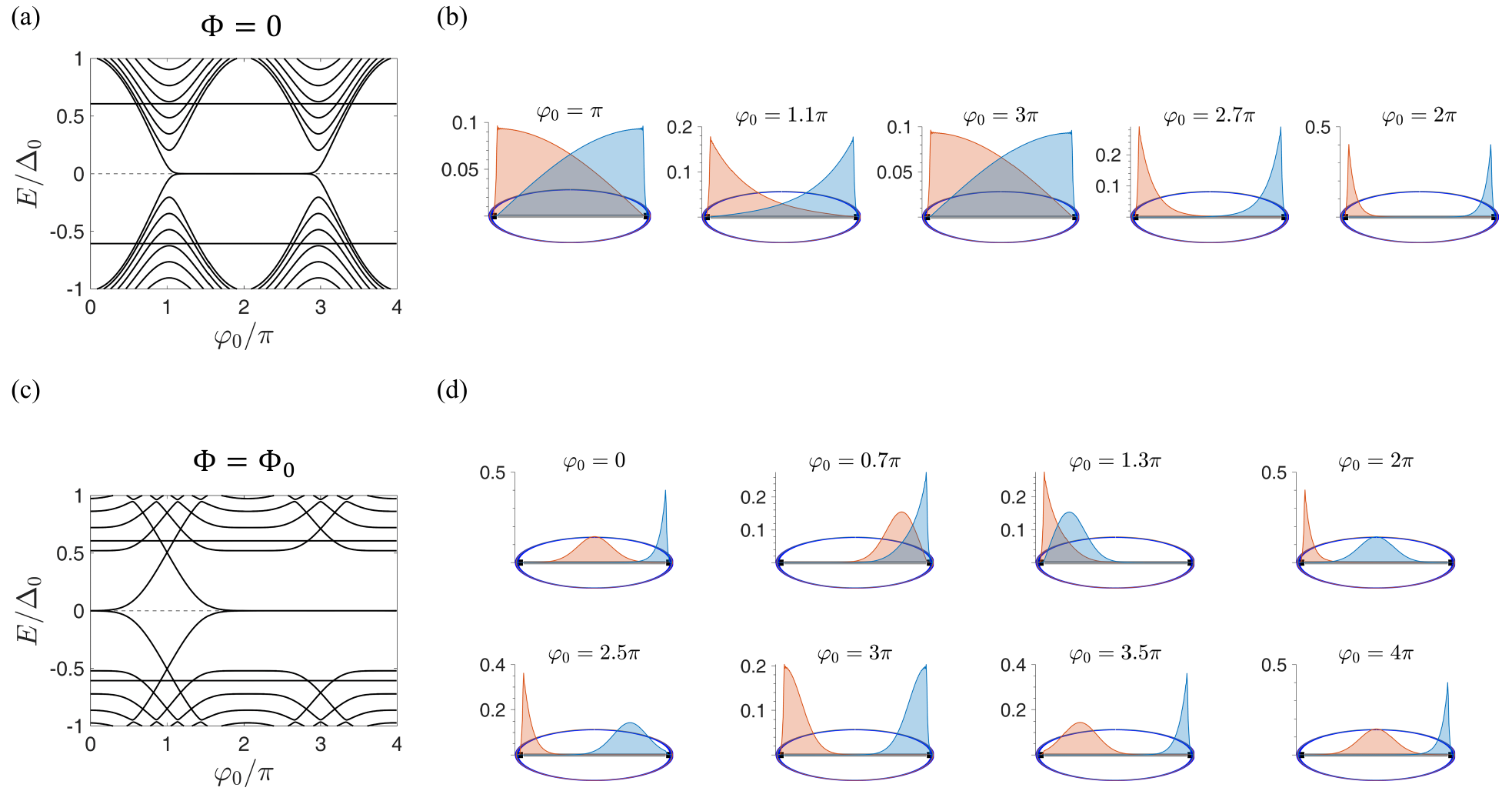}
    \caption{(a) The energy spectrum as a function of the phase difference $\varphi_0$ for $\Phi=0$ in the absence of the junction-boundary coupling and (b) the spatial distributions of wave functions of the Majorana modes composing the energy excitation around $E=0$ in (a) for $\pi\leq \varphi_0 \leq 3\pi$. In (b), the gray horizontal line segment and the blue circle depict the JJ and the boundary, respectively, and the vertical axis measures the local amplitudes of the wave functions. The black squares represent the joint between the junction and the boundary. (b) The well-separated Majorana zero modes (MZMs) at the ends of the junction at $\varphi_0 = 2\pi$ and the increase of the wave function overlap through the junction for $\varphi_0$ away from $2\pi$. Pannels (c) and (d) are the same plots as in (a) and (b) but for $\Phi= \Phi_0$. In (d), one MZM is localized at the Josephson vortex, which moves as $\varphi_0$ changes. The other MZM is bound at one of the junction ends. As the mobile mode approaches the end mode, they are hybridized [the upper panel of (d)]. When the mobile mode approaches the empty end, it then becomes an MZM bound to the end [lower panel of (d)] since there has to be a pair of MZMs and there is no way for MZMs to leak out of the junction.
    The computation is performed with the parameters of $L_J=1.5\mu$m, $L_b=9\mu $m, $v_b=v_F$, $v_J\approx 0.038 v_F$, and $\Delta_0=151\mu$eV.}
    \label{fig: level and wave function}
\end{figure*}

To see the boundary effect in more detail, we introduce an effective 1D model of the JJ device by treating conducting channels along the boundary of the SC as 1D gapless modes, as shown in Fig. \ref{fig: 1d model}. In the short junction limit $d\ll \xi=\hbar v_F/\Delta_0$, the Hamiltonian can be reduced to a low-energy 1D Hamiltonian described by two counter propagating Majorana fields $\eta_{L,R}(s)$ satisfying $\eta_{L,R}^{\dag}=\eta_{L,R}$,
\begin{equation}\label{eq: Effect 1D total H}
    H=\frac{1}{2}\int ds (\eta_{R},\eta_{L})^{T}\mathcal{H}_{tot}
    \begin{pmatrix}
        \eta_{R} \\ 
        \eta_{L}
    \end{pmatrix}.
\end{equation}
$\mathcal{H}_{tot}$ consists of three contributions, $\mathcal{H}_{tot}=\mathcal{H}_{J}+\mathcal{H}_{b}+\mathcal{H}_{c}$. $\mathcal{H}_{J}=\left[-i\hbar v_{J}\sigma_{z}\partial_{s}+m(s,\Phi,\varphi_0)\sigma_{y}\right]\Theta(s)\Theta(L_J-s)$ describes the counter propagating Majorana fields inside the junction region (see the black arrows of the gray region in Fig. \ref{fig: 1d model}) with velocity $v_J$ \cite{Fu_and_Kane_PRL2008, Potter_and_Fu_PRB2013}. Coupling between the two fields is given by $m(s,\Phi,\varphi_0)=\Delta_{0}\cos\frac{\varphi(s,\Phi,\varphi_0)}{2}$ where the gauge-invariant phase difference is 
\begin{equation}\label{eq: Gauge invariant phase difference}
    \varphi(s,\Phi,\varphi_0)=\frac{2\pi \Phi}{L_J \Phi_0}s-\varphi_0\,.
\end{equation}  
Here $\varphi_0$ is the superconducting phase difference between the two SCs, $\Phi=B_0 L_J(d+2\lambda_L)$ and $\Phi_0=h/2e$ is the flux quantum.
$\mathcal{H}_b=\left[-i\hbar v_b \sigma_{z}\partial_{s}\right](\Theta(-s)+\Theta(s-L_J))$ 
describes the two counter-propagating Majorana boundary modes, 
depicted by the blue arcs in Fig. \ref{fig: 1d model}. In 3D TI JJ system, these modes 
appear along the boundary of the SCs when the region outside the proximity-induced superconductors is gapped out by a magnetic field.
They can also appear when the outside region is ballistic. There are in general many surface modes in the outside region, but among them, only one Majorana mode couples with the Majorana channel inside the junction at each junction end, as shown in Refs. \cite{Buttiker_PRB2012,Wieder_PRB2013}. The two Majorana modes, one at each junction end, can form the chiral boundary modes described by $\mathcal{H}_b$.
The couplings between the outer boundary fields and the inner junction fields (see the right dashed circle in Fig. \ref{fig: 1d model}) are modeled by delta energy barriers at the two ends of the junction $\mathcal{H}_{c}=\left(U\delta(s)+U\delta(s-L_J)\right)\sigma_{y}$. This coupling results in the scattering matrix at the ends \cite{Wieder_PRB2013}:
\begin{equation}\label{eq: S matrix point scatterer}
    \begin{array}{ll}
    \displaystyle{r=\tanh \frac{U}{\hbar (v_J+v_b)/2}},&\displaystyle{t'=\frac{1}{\cosh\frac{U}{\hbar (v_J+v_b)/2}}},\\
    \\
    \displaystyle{t=\frac{1}{\cosh\frac{U}{\hbar (v_J+v_b)/2}}}, & \displaystyle{r'=-\tanh \frac{U}{\hbar (v_J+v_b)/2}}.
\end{array}
\end{equation}
For $U\rightarrow\infty$, $R=|r|^2=1$ and the junction are totally decoupled from the boundary, while for $U\rightarrow0$, $R=0$, meaning that the junction and the boundary are fully connected.
Since the fields $\eta_{R,L}(s)$ originate from the TI surface states, they get Berry phase $\pi$ for a round trip along the boundary(due to the momentum-spin locking of Dirac fermions), hence satisfying the antiperiodic boundary condition $\eta_{R,L}(s+L_J+L_b/2)=-\eta_{R,L}(s)$. Note that the scattering Hamiltonian $\mathcal{H}_c$ allows us to tune the coupling between the boundary and the junction, and, thus, renders a systematic analysis of the boundary effect.  

\begin{figure*}
\centering
\includegraphics[width=0.95\textwidth]{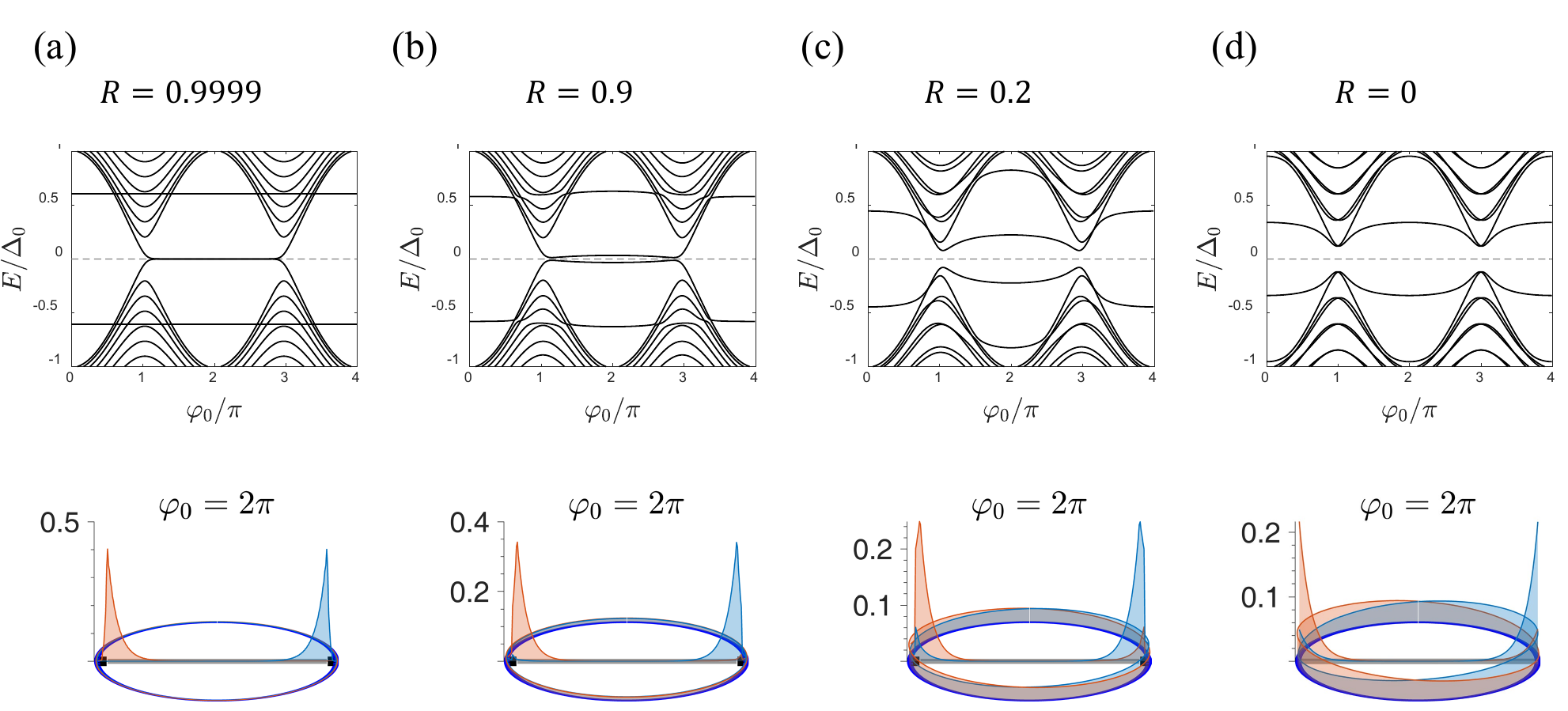}
\caption{Evolution with the junction-boundary coupling from $R=0.9999$ (decoupling) to $R=0$ (strong coupling) at $\Phi=0$. Upper panels: The energy splitting of the Majorana modes around $\varphi_0=2\pi$ increases as $R$ decreases, exhibiting the the change of the periodicity from $4\pi$ [i.e., $E_{GS}(\varphi_0+4\pi)=E_{GS}(\varphi_0)$] at $R=0.9999$ to $2\pi$ [$E_{GS}(\varphi_0+2\pi)=E_{GS}(\varphi_0)$] at $R=0$. Lower panels: The spatial distributions of the Majorana wave functions at $\varphi_0=2\pi$ for the different values of the junction-boundary coupling. As the energy splitting increases, the amplitude (the height along the circle) of the boundary wave functions outside the junction increases, while the direct overlap between the Majorana modes inside JJ remains zero, indicating the boundary-induced energy splitting.  
}
\label{fig: Boundary effect at zero B}
\end{figure*}

We calculate the energy spectrum and wave functions by discretizing the continuum Hamiltonian (\ref{eq: Effect 1D total H}) on a lattice and diagonalizing the discretized Hamiltonian with the antiperiodic boundary condition \cite{ZiXiang_FDP_ScienceAdvances2018,Drell_FDP_PRD1976}.
The parameters $L_J=1.5\mu$m, $L_b=9\mu $m, and $\Delta_0=151\mu$eV are used. 
Lattice constant $a$ is set to be 6.58 nm, so that the Dirac cone spectrum of 1D channels is realized within the energy range $30 \Delta_0$. Furthermore, to model $\mathcal{H}_c$ we added 10 lattice sites between the boundary and the junction (region represented by the dark box in Fig. \ref{fig: 1d model}) and give the potential $(U/(10a))\sigma_{y} $ to that region. Note that the potential goes to delta function when $a\rightarrow 0$.
The velocity of the boundary modes is set to be $v_b=v_F=4\times 10^{5}$m/s, and the velocity of the junction is given by $v_J\sim v_F\Delta_0^2/(\mu_s^2 +\Delta_0^2)$, where $\mu_s$ is the chemical potential in the superconducting region.
Due to the renormalization of the chemical potential induced by the proximity effect from the SC, $\mu_s$ is lifted from the original value of TI surface mode in the absence of the SCs \cite{Stanescu_PRB2010,Lababidi_Mahmoud_and_Zhao_PRB2011}. According to numerical simulations \cite{Lababidi_Mahmoud_and_Zhao_PRB2011}, $\mu_s$ increases to $4 \Delta_0 \sim 5\Delta_0$ even when the chemical potential is set to the Dirac point in the absence of the SCs. 
Based on this, we analyze the system with $\mu_s=5\Delta_0=0.755 \mu$eV, hence $v_J\approx 0.038\,v_F\approx 1.52\times 10^{4}$m/s.

\subsection{Majorana zero modes in the junction}

Let us first consider the case where the junction is fully decoupled from the boundary by setting $U\rightarrow\infty$. 
From Eq. \eqref{eq: S matrix point scatterer}, $r'\rightarrow -1$ and $r\rightarrow 1$. The quantization condition of the boundary mode is $2kl+\pi =2\pi n$, which corresponds to the boundary channels with $L_b=2l$ in Sec. \ref{sec2}. This energy level is not affected by $\varphi_0$, resulting in flat energy level as shown in Figs. \ref{fig: level and wave function}(a) and \ref{fig: level and wave function}(c). 

Figure. \ref{fig: level and wave function}(a) shows the energy spectrum of the junction at zero magnetic field ($\Phi=0$). In this case, the Josephson coupling $m=\Delta_0\cos\frac{\varphi_0}{2}$ is constant along the junction, with the spectrum gapped by $2|m|$. As $\varphi_0$ changes from $\varphi_0<\pi$ to $\varphi_0>\pi$, the gap is closed and inverted from $m>0$ to $m<0$, entering into the topological phase where two Majorana end modes appear as
\begin{align}
    \gamma_{1}=&\int_{0}^{L_J}ds \frac{e^{m s/\hbar v_J}}{\sqrt{2\hbar v_J/ |m|}}\left(\eta_{R}(s)-\eta_{L}(s)\right), \label{eq: gamma1-end-zero-B}\\ 
    \gamma_{2}=&\int_{0}^{L_J}ds\frac{e^{-m (s-L_J)/\hbar v_J}}{\sqrt{2\hbar v_J/|m|}}\left(\eta_{R}(s)+\eta_{L}(s)\right).\label{eq: gamma2-end-zero-B}
\end{align}
 The evolution of the Majorana wave functions with the phase $\varphi_0$ is shown in Fig. \ref{fig: level and wave function}(b).
The pair of energy bands of the Majorana modes lying near the zero energy yields a flatlike feature with $m<0$ in the window of $\pi < \varphi_0 < 3 \pi$, showing $4\pi$-periodicity.
This kind of double period attributed to the parity degree of freedom has been discussed in many other setups both theoretically
and experimentally.

\begin{figure*}
\centering
\includegraphics[width=0.95\textwidth]{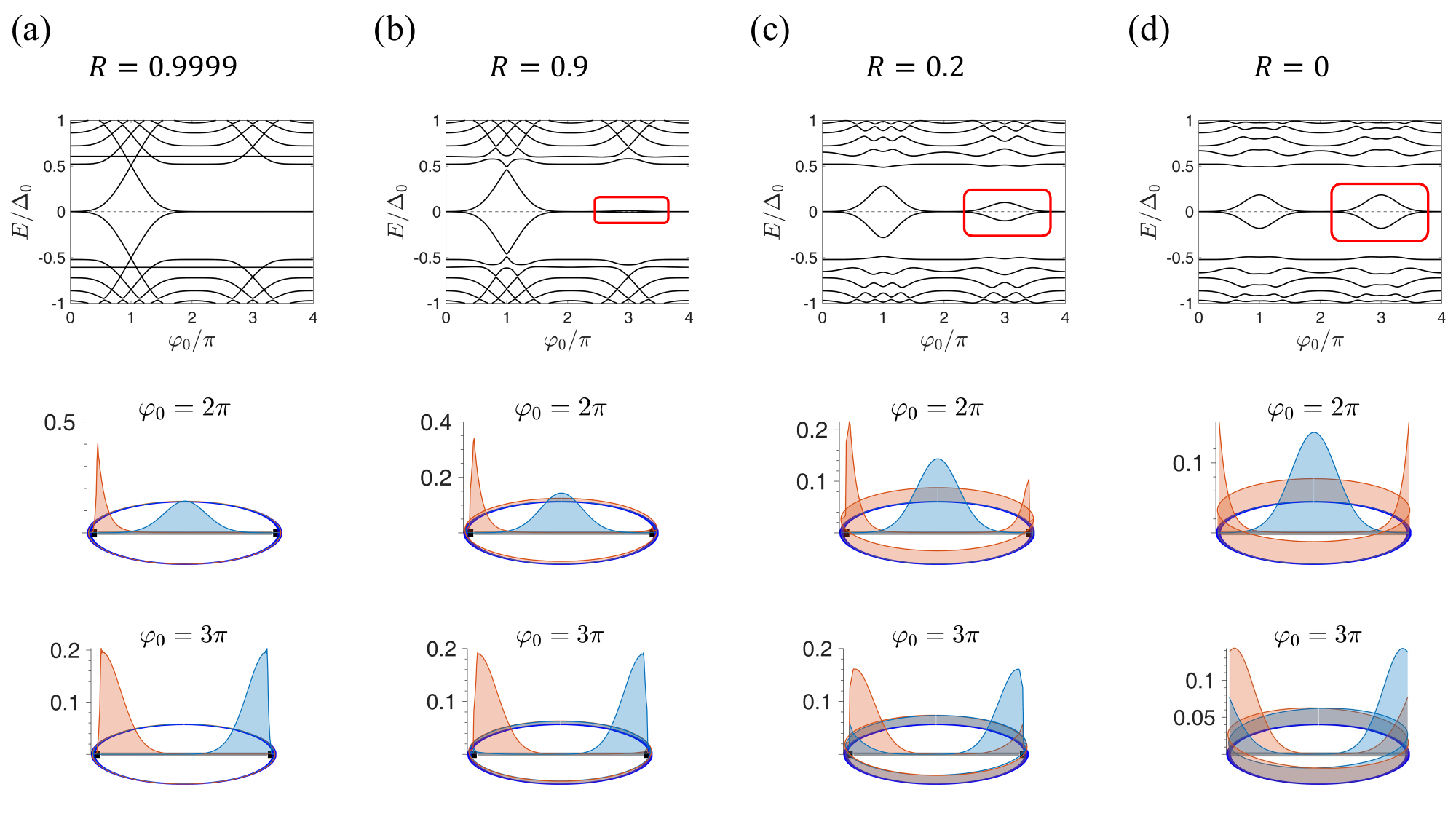}
\caption{Same plots as in Fig.~\ref{fig: Boundary effect at zero B}, but for $\Phi = \Phi_0$. 
Top panels: The ground-state energy exhibits $4\pi$ periodicity as a function of $\varphi_0$ at $R=0.9999$ in (a) due to the flatlike bands of the MZMs around $E=0$. It becomes $2\pi$-periodic in the energy spectrum as $R$ decreases to $R=0$ in (d) due to the energy splitting at $\varphi_0=3\pi$. Middle panels: At $\varphi_0 = 2 \pi$, one MZM is localized in the Josephson vortex at the center of the JJ, while the other is localized at the left end of the JJ. The latter becomes delocalized along the boundary as $R$ decreases to 0. The MZMs have negligible spatial overlap, resulting in zero energy states.  
Bottom panels: At $\varphi_0 = 3 \pi$, the Josephson vortex is located at the right end of the JJ so that there is one Majorana mode at each end of the JJ. The modes are MZMs when $R=1$, as they have no spatial overlap inside the JJ. As $R$ decreases, the modes are more coupled though the boundary channels, resulting in a larger energy splitting near $\varphi_0 = 3 \pi$ (marked as red boxes in the top panels).
}
\label{fig: Boundary effect at single flux}
\end{figure*}

For the magnetic field near the single flux quantum $\Phi\sim\Phi_0$ applied to the junction, the energy spectrum and the wave functions are plotted in Figs. \ref{fig: level and wave function}(c) and \ref{fig: level and wave function}(d). In this case, the variation of $\varphi(s,\Phi,\varphi_0)$ along the junction causes the variation of Josephson coupling $m(s,\Phi,\varphi_0)$. Then a MZM localized at the Josephson vortex $\varphi(s,\Phi,\varphi_0)=\pi$ (equivalently, $m(s,\Phi,\varphi_0)=0$) appears as 
\begin{equation}\label{eq: gamma-mobile-B}
    \gamma_{1}=\int_{0}^{L_J}ds\frac{e^{-\frac{(s-s_1)^2}{2\lambda_M^2}}}{\sqrt{2\pi\lambda_M^2}}\frac{\left(\eta_{R}(s)-\eta_{L}(s)\right)}{\sqrt{2}},
\end{equation}
where $\lambda_{M}=\sqrt{\hbar v_J L_J/(2\pi\Phi/\Phi_0)}$, $s_{1}=(\varphi_0/2\pi -1/2)L_J/(\Phi/\Phi_0)$. This localized vortex mode is a mobile Majorana mode whose position is tuned by the global phase difference $\varphi_0$ as shown in Fig.. \ref{fig: level and wave function}(d) and whose width is controlled by the magnetic field. \cite{Potter_and_Fu_PRB2013}.
The other Majorana mode, which pairwise appears together with the mobile Majorana $\gamma_1$, is the end mode localized at the junction end of $m(s,\Phi,\varphi_0)<0$, in this case, at the right end,
\begin{equation}\label{eq: gamma-end-B}
    \gamma_{2}=\int_{0}^{L_J}ds e^{-\int_{L_J}^{s}ds'\frac{m(s',\Phi,\varphi_0)}{\hbar v}}\frac{\eta_{R}(s)+\eta_{L}(s)}{\sqrt{2}}\, .
\end{equation}
The energy of the fermion states formed by $\gamma_1$ and $\gamma_2$ is split as the mobile mode $\gamma_1$ approaches and hybridizes with the end mode $\gamma_2$ by change of $\varphi_0$. On the other hand, when the mobile Majorana mode moves outward from the end mode and approaches the opposite end of the junction, it becomes another end mode localized at the opposite end. In this case $m(s,\Phi,\varphi_0)<0$ all through the junction. This behavior results in $4\pi$-periodic energy levels.


\subsection{Boundary induced energy level splitting}\label{sec3.2}

The case when the boundary mode couples with the junction is shown in Fig. \ref{fig: Boundary effect at zero B}. 
When the coupling turns on, the bound states are now formed by the junction mode and boundary mode.
When $R=1$, $U\rightarrow\infty$, the chiral channels and junction channels are totally decoupled. The energy levels are shown as one level made by the boundary channels and the other levels made by the Majorana modes inside the junction as discussed in the previous subsection. 
When $R$ slightly decreases from $1$, the junction channels weakly couple with the boundary channels. As shown in Fig. \ref{fig: Boundary effect at zero B}, the two Majorana modes hybridize not only through the wave function tails inside the junction, but also through the boundary channels.
Note that the Majorana modes at the two ends of the junction at $\varphi_0 = 2 \pi$ (see the bottom panels of Fig. \ref{fig: Boundary effect at zero B}) have no overlap inside the junction, when the decay length of the Majorana modes, $l_\textrm{decay} = \hbar v_J / |m|$, is much shorter than the junction width $L_J$. Even in this case, the Majorana modes have overlap through the boundary mode, when there is coupling between the junction and the boundary mode.
This boundary-mediated overlap results in the splitting of the energy levels formed by the Majorana modes at the topological phase $m < 0$, as discussed in Sec. \ref{sec2}.
The energy splitting weakens the $4\pi$ periodicity of the energy levels in the sense that as $R$ becomes smaller, the two end Majorana modes (one at each junction end) are more hybridized through the boundary mode outside the junction, resulting in more energy splitting of the states formed by the end Majorana modes.

The $4 \pi$ periodicity is weakened also when the magnetic flux of $\Phi \simeq \Phi_0$ near the flux quantum is applied to the junction (see Fig. \ref{fig: Boundary effect at single flux}). 
At this magnetic flux, there is one mobile Majorana mode localized at the Josephson vortex (namely at $\varphi (s, \Phi, \varphi_0) = \pi$; see Eq. (\ref{eq: Gauge invariant phase difference})) inside the junction and the other Majorana mode localized at one end of the junction. 
The two modes have spatial overlap inside the junction around $\varphi_0 = \pi$, resulting in no zero energy state in the energy spectrum (independently of $R$; see the top panels of Fig. \ref{fig: Boundary effect at single flux}). As $\varphi_0$ increases away from $\varphi_0 = \pi$, their overlap inside the junction becomes reduced, as they become separated from each other. Around $\varphi_0 = 2 \pi$, the overlap vanishes (in the parameters chosen for Fig. \ref{fig: Boundary effect at single flux}) and the two modes become MZMs. The overlap remains negligible (hence the two modes remain as MZMs), even in the presence of the coupling to the boundary channels with $R < 1$ (the middle panels of Fig. \ref{fig: Boundary effect at single flux}). As $\varphi_0$ further increases to $3 \pi$, the Majorana mode localized at the Josephson vortex approaches the junction end opposite to the other end where the other Majorana mode is localized. While the two modes do not have overlap inside the junction, they become hybridized through the boundary channels in the presence of the coupling to the boundary channels with $R < 1$ (the bottom panels of Fig. \ref{fig: Boundary effect at single flux}). As a result, the $4 \pi$ periodicity is weaken as the coupling becomes stronger (the top panels of Fig. \ref{fig: Boundary effect at single flux}).

\section{Fraunhofer pattern}\label{sec4}
In this section, we discuss the effect of the boundary channels on the magnetic field dependence of the Josephson critical currents, i.e., Fraunhofer patterns. 
In a non-topological JJ with a sinusoidal current-phase relation $I_J(\varphi)=I_c\sin\varphi$, the critical current exhibits a diffraction-like feature due to the interference of the Josephson current density along the junction under the influence of the $B$-field. The critical current drops to zero at nodes, where the magnetic flux $\Phi$ threading the junction area equals an integer multiple of the flux quantum $\Phi_0$, $\Phi = \nu \Phi_0$ ($\nu = 1,2,3, \cdots$), due to destructive interference. 
We will show below that the boundary effects in the topological JJ with Majorana modes induce changes at the nodes of the Fraunhofer pattern, distinguishing it from the non-topological one.

The Fraunhofer pattern is obtained by calculating the Josephson current at zero temperature limit \cite{Beenakker_FDP_GEP_SciPostPhys2021}, 
\begin{equation}\label{eq: Josephson current}
    I_J=\frac{2e}{\hbar}\frac{\partial E_{GS}}{\partial \varphi_0}
\end{equation}
where the ground state energy $E_{GS}(\Phi, \varphi_0)$ is given by the sum of all negative energy levels $E_n(\Phi,\varphi_0)$ of the Hamiltonian in Eq. \eqref{eq: Effect 1D total H},  
\begin{equation}\label{eq: gs energy}
    E_{GS}(\Phi,\varphi_0)=\frac{1}{2}\sum_{E_n<0} E_n(\Phi,\varphi_0).
\end{equation}
The critical current $I_c$ is then obtained as the maximum Josephson current $I_J$.  
\begin{figure}
    \centering
    \includegraphics[width=0.49\textwidth]{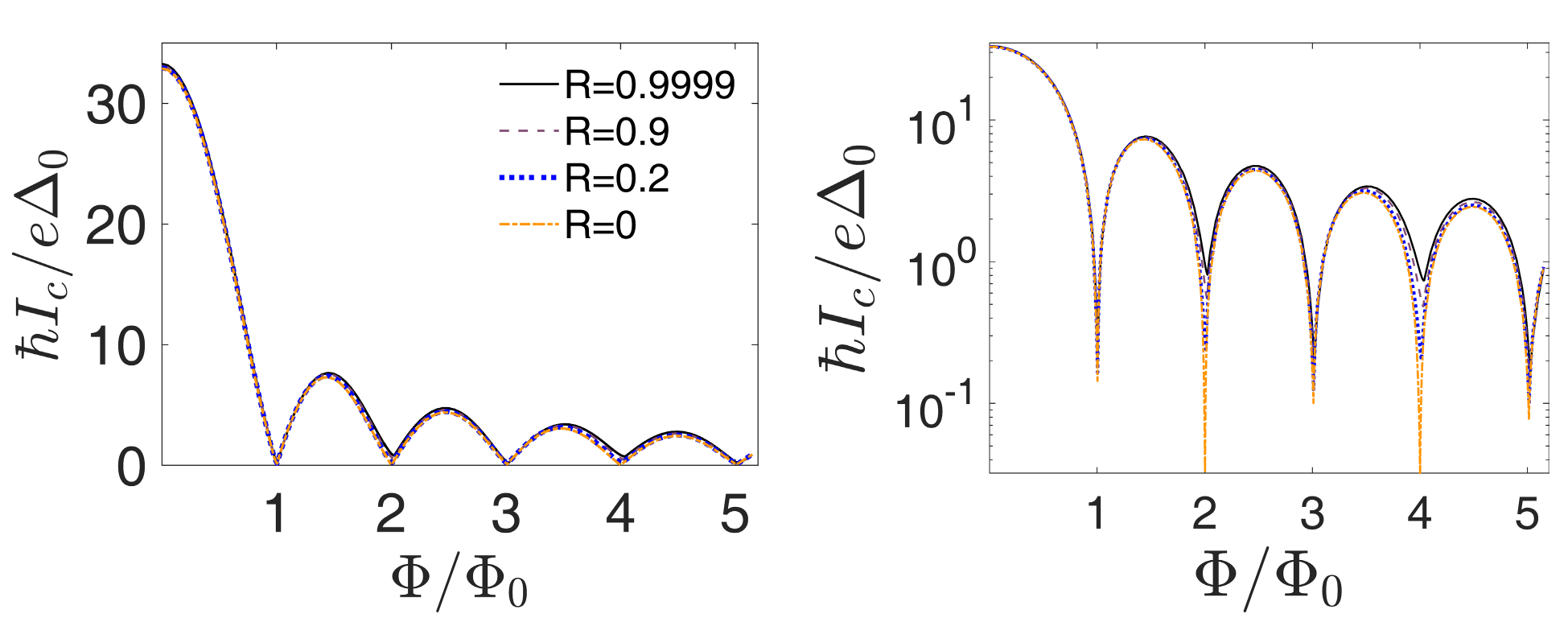}
    \caption{Fraunhofer pattern. Left: Critical currents as a function of the magnetic flux $\Phi/\Phi_0$. Right: The same plot in the logarithmic scale. The values of $R = 0.9999$, 0.9, 0.2, 0 are chosen. 
    The other parameters are
    $L_J=2.5\mu$m, $\Delta_0=151\mu eV$, $L_b=9\mu$m, $v_b=v_F$, and $\mu_s=10\Delta_0=1.51$meV, where $ v_J\approx 0.0099v_F=3.96\times 10^{3}$m/s.}
    \label{fig: fraunhofer plot}
\end{figure}
The ground state energy in Eq. \eqref{eq: gs energy} consists of the energy contributions from the MZMs and energy levels of the trivial mid-gap states.
As discussed before, application of the integer number $\nu$ of the flux quanta $\Phi=\nu\Phi_0$ creates the $\nu$ number of MZMs in the JJ, and $\varphi_0$ determines the positions of the MZMs according to Eq. \eqref{eq: Gauge invariant phase difference}.

We compute the Fraunhofer patterns in Fig. \ref{fig: fraunhofer plot}. The notable feature of the resulting Fraunhofer patterns is the nonvanishing currents that lift the nodes at $\Phi = \nu\Phi_0$. 
At these nodes, the finite critical current is determined by the energy splitting due to hybridizations of the MZMs which occur near the junction end (discussed later), while the non-Majorana contribution to the current vanishes (as in the non-topological JJs). 
Since the current flows at the junction end where the hybridization takes place, it is sensitive to the junction-boundary coupling. 

In the case of strong junction-boundary coupling ($R= 0$), there appears an even-odd behavior. 
At $\Phi=\nu\Phi_0$ with odd $\nu$, the nodes of the Fraunhofer patterns are lifted, while the node lifting is negligible at even $\nu$
(orange dash-dotted in Fig. \ref{fig: fraunhofer plot}).
This is understood as followings.
There appears an extended MZM
along the boundary channels, depending on the even-oddness of $\nu$ .
When $\nu$ is odd, there appear $\nu$ MZMs in the JJ and an extended MZM along the boundary,
maintaining that the total number of MZMs in the system is even [see the middle panel of Fig. \ref{fig: Boundary effect at single flux}(d) for $\nu=1$]. 
Then hybridization happens between an MZM close to the Junction end and the extended MZM, while the MZM of the JJ approaches to a junction end as $\varphi_0$ varies.
This hybridization results in the energy splitting, which produces the finite currents at the junction end. 
This is the origin of the node lifting.
In contrast, for even $\nu$, where there are even number of MZMs in JJ, there are no extended MZMs along the boundary \cite{choi2019josephson} (maintaining that the total number of MZMs is even).
As the result, no hybridization happens, and node liftings are negligible. 


On the other hand, in the case of weak junction-boundary coupling ($R= 1$), another behavior happens. Node lifting occurs at all the integer $\nu$ (dark solid line in Fig. \ref{fig: fraunhofer plot}). 
Contrary to the $R=0$ case, the MZMs bound to the junction ends appear at all integer $\nu$ (one end MZM at a junction end for odd $\nu$, and two MZMs at both junction ends for even $\nu$). Then hybridization happens between one of the end MZMs and a mobile MZM, depending on $\varphi_0$, resulting in a nonvanishing critical current at the junction end [see Figs.~\ref{fig: level and wave function}(c) and \ref{fig: level and wave function}(d) for $\nu=1$]. This explains the node lifting at all integer values of $\nu$.

We note that effects of MZMs on Fraunhofer patterns have been discussed in Ref.~\cite{Potter_and_Fu_PRB2013}. 
In Ref.~\cite{Potter_and_Fu_PRB2013}, hybridization occurs between an MZM on the top surface and another on the bottom surface, differently from our study (where the hybridization happens on a top surface).
Recently there was a report on experimental detection~\cite{Fraunhofer_GuangYue_PRB2024} of Fraunhofer patterns in Bi$_2$Se$_3$-Nb TI JJ, which shows a similar behavior to the $R=0$ case in our study.

\section{Summary}\label{conc}
We discussed the stability of MZMs in a planar geometry of a topological JJ, focusing on their coupling with boundary modes along the boundary of the junction device. Our model takes account of tunnel coupling at the ends of the junction with the 1D chiral boundary channel, resulting in the scattering matrix connecting Majorana modes inside the junction and outer boundary modes. It also includes the effects of the superconducting phase difference $\varphi_0$ and the magnetic flux $\Phi$. By solving this effective 1D scattering model, we calculate the boundary effect on the energy spectrum and Josephson effects of the topological JJ for the coupling strength from the decoupled ($R$=1) to coupled ($R$=0) regime. For zero magnetic flux $\Phi$=0, the coupling with the boundary modes induces the effective coupling between MZMs inside the junction, which lifts the degeneracy at zero energy in the range of $\pi<\varphi_0<3\pi$. We found the form of the energy splitting analytically, showing its independence on the loop length of the boundary, in the weak coupling limit, and also the evolution of the energy spectrum from $4\pi$- to $2\pi$-periodicity with changing $R$. For the presence of a flux quantum $\Phi=\Phi_0$ within the junction, the degeneracy of the Majorana modes around $\varphi_0 = 3\pi$ is broken due to the boundary-induced coupling. However, we found that the MZMs at $\varphi_0 = 2\pi$ remain at zero energy due to the fact that a single Majorana mode in the junction is localized at the center of the junction and is well separated from the boundary. We finally calculated critical currents as a function of $\Phi$ (the Fraunhofer pattern) for different junction-boundary coupling strengths. Differently from conventional JJs, we found nonzero currents (Fraunhofer pattern node lifting) at certain integer flux quanta $\Phi = \nu \Phi_0$, induced by the hybridization between MZMs in the system, which is sensitive to the junction-boundary coupling.

\begin{acknowledgments}
    This work is supported by the Korea NRF (SRC Center for Quantum Coherence in Condensed Matter, Grants No. RS-2023-00207732 and No. 2023R1A2C2003430). S.P. acknowledges the financial support from the Institute for Basic Science (IBS) in the Republic of Korea through the Project No. IBS-R024-Y4.
\end{acknowledgments}
\appendix
\section{Derivation of Eq. (\ref{eq: gamma1-end-zero-B})-(\ref{eq: gamma-end-B})}\label{appendix: derivation of formula}
For derivation of Eqs. (\ref{eq: gamma1-end-zero-B})-(\ref{eq: gamma-end-B}), we first diagonalize the Hamiltonian in Eq. (\ref{eq: Effect 1D total H}) so that
\begin{align}\label{eq_appendix: H-diagonalized}
    H=&\frac{1}{2}\int ds (\eta_R(s),\eta_L(s))\mathcal{H}_{tot}
    \begin{pmatrix}
    \eta_R(s)\\ \eta_L(s)
    \end{pmatrix}
    \nonumber\\ 
    =&
    \sum_{n=0}^{\infty}E_n\left(a_n^{\dag}a_{n}-\frac{1}{2}\right)\,,
\end{align}
where $a_{n}$ is the fermion annihilation operator for the excitation of n-th energy level with an energy $E_n\geq 0$, and $\mathcal{H}_{tot}$ is given by (with $\epsilon\rightarrow 0$) 
\begin{displaymath}
    \mathcal{H}_{tot}
    =
    \begin{cases}
        -i\hbar v_J\sigma_{z}\partial_{s}+m(s,\Phi,\varphi_0) \sigma_{y}&(0<s<L_J)\\ 
        -i\hbar v_b \sigma_{z}\partial_{s} & (s>L_J, s<0)\\ 
        -i\hbar\sqrt{v}\partial_{s}\sqrt{v}\sigma_{z}+U\sigma_{y}\delta(s)&(|s|<\epsilon)\\ 
        -i\hbar\sqrt{v}\partial_{s}\sqrt{v}\sigma_{z}+U\sigma_{y}\delta(s-L_J)&(|L_J-s|<\epsilon).
    \end{cases}
\end{displaymath}
Here the JJ region is $0<s<L_J$, and the boundary regions are $s>L_J$ and $s<0$. In $|s|<\epsilon$, $|L_J-s|<\epsilon$ the energy barrier $U/\epsilon$, which becomes delta function for $\epsilon\rightarrow 0$, are given to impose the coupling between JJ and the boundary. $v(s)$ is the position-dependent velocity which is smoothly varying from $v_J$ to $v_b$ within this barrier region. 
We then find the MZMs $\gamma_{1,2}$ constituting the zero energy excitation $a_0$ by the relation $a_0=(\gamma_1+i\gamma_2)/\sqrt{2}$, or equivalently  $\gamma_1=(a_0+a_0^{\dag})/\sqrt{2}$, $\gamma_2=-i(a_0-a_0^{\dag})/\sqrt{2}$.

The diagonalization in Eq. (\ref{eq_appendix: H-diagonalized}) is acheived by expressing $(\eta_R(s), \eta_L(s))^{T}$ into
\begin{equation}\label{eq_appendix: Majorana field-span by excitation}
    \begin{pmatrix}
        \eta_{R}(s)\\ \eta_{L}(s)
    \end{pmatrix}
    =
    \sum_{n=0}^{\infty}
    \left[
    \begin{pmatrix}
        u_{R,n}(s)\\ u_{L,n}(s)
    \end{pmatrix}
    a_n
    +
    \begin{pmatrix}
        u_{R,n}^{\ast}(s)\\ u_{L,n}^{\ast}(s)
    \end{pmatrix}
    a_n^{\dag}
    \right], 
\end{equation}
where $u_{R,n}(s)$, $u_{L,n}(s)$ are complex functions of $s$ which satisfy the eigenvalue equation $\displaystyle{\mathcal{H}_{tot}(u_{R,n},u_{L,n})^{T}=E_n(u_{R,n},u_{L,n})^{T}}$.
When the junction and the boundary are decoupled by the condition $U\rightarrow\infty$, the scattering matrix in Eq. (\ref{eq: S matrix point scatterer}) becomes $r=1$, $r'=-1$, which provide the boundary conditions 
for the $(u_{R,n},u_{L,n})^T$ inside the junction,
\begin{equation}\label{eq_appendix: BC}
   \begin{array}{rl}
    u_{R,n}(0^{+})=& -u_{L,n}(0^{+})\,,\\
    u_{R,n}(L_J-0^{+})=&u_{L,n}(L_J-0^{+})\, .
   \end{array}
\end{equation}
The zero energy solution in the junction region satisfies 
\begin{equation}\label{eq_appendix: zero E equation}
    \left(-i\hbar v_J\sigma_{z}\partial_{s}+m(s,\Phi,\varphi_0)\sigma_{y}\right)
    \begin{pmatrix}
        u_{R,0}(s)\\ u_{L,0}(s)
    \end{pmatrix}
    =0.
\end{equation} 

The solution basis states with zero energy are then 
\begin{displaymath}
    \frac{1}{\sqrt{2}}\begin{pmatrix}
        1\\ \pm 1
    \end{pmatrix}
    e^{\mp \int^{s}ds'\frac{m(s',\Phi,\varphi_0)}{\hbar v_J}}.
\end{displaymath}
The zero enery solutions $(u_{R,0}(s),u_{L,0}(s))^{T}$ are superpositions of these basis states which satisfy the boundary condition Eq. (\ref{eq_appendix: BC}). 

The zero energy excitation $a_0$ satisfying the relation Eq. (\ref{eq_appendix: Majorana field-span by excitation}) is then given by 
\begin{displaymath}
    a_0=\int ds (u_{R,0}^\ast(s),u_{L,0}^\ast(s))\begin{pmatrix}
        \eta_R(s)\\ \eta_L(s)
    \end{pmatrix}.
\end{displaymath}
The MZMs constituting $a_0$ given by $\gamma_1=(a_0+a_0^{\dag})/\sqrt{2}$, $\gamma_2=-i(a_0-a_0^{\dag})/\sqrt{2}$ are
\begin{align}
    \gamma_1=&\frac{1}{\sqrt{2}}\int ds (\text{Re}(u_{R,0}(s)),\text{Re}(u_{L,0}(s)))
    \begin{pmatrix}
        \eta_R(s)\\ \eta_L(s)
    \end{pmatrix},\\
    \gamma_2=& -\frac{1}{\sqrt{2}}\int ds (\text{Im}(u_{R,0}(s)),\text{Im}(u_{L,0}(s)))
    \begin{pmatrix}
        \eta_R(s)\\ \eta_L(s)
    \end{pmatrix}.
\end{align}
Note that $(u_{R,0}^{\ast}(s),u_{L,0}^{\ast}(s))^T$ also satisfies Eqs. (\ref{eq_appendix: zero E equation}) and (\ref{eq_appendix: BC}). 
As $u_{L(R),0}\pm u_{L(R),0}^{\ast}$ satisfies Eqs. (\ref{eq_appendix: zero E equation}) and (\ref{eq_appendix: BC}), the real-valued solutions of Eq. (\ref{eq_appendix: zero E equation}) satisfying the boundary conditions in Eq. (\ref{eq_appendix: BC}) provide the MZMs composing the zero energy excitation $a_0$.

If $m(s,\Phi,\varphi_0)<0$ all inside the junction, the wave functions
\begin{align}
    \begin{pmatrix}
        \text{Re}(u_{R,0}(s))\\\text{Re}(u_{L,0}(s))
    \end{pmatrix}
    =&\frac{\mathcal{N}_1 }{\sqrt{2}}\begin{pmatrix}
        1\\ -1
    \end{pmatrix}
    e^{\int_{0}^{s}ds'\frac{m(s')}{\hbar v_J}}\, ,
    \nonumber\\ 
    \begin{pmatrix}
        \text{Im}(u_{R,0}(s))\\\text{Im}(u_{L,0}(s))
    \end{pmatrix}
    =&\frac{\mathcal{N}_2}{\sqrt{2}} \begin{pmatrix}
        1\\ 1
    \end{pmatrix}
    e^{-\int_{L_J}^{s}ds'\frac{m(s')}{\hbar v_J}}
    \nonumber
\end{align}
with proper real-valued normalization constant $\mathcal{N}_{1,2}$ satisfy the boundary condition in Eq. (\ref{eq_appendix: BC}) of the system when $\hbar v_J|m|\ll L_J$, since both wave functions exponentially decay into the junction from each end $s=0$, $s= L_J$. The corresponding MZMs are given by  
\begin{align}
    \gamma_1 =& \mathcal{N}_1\int_{0}^{L_J} ds e^{\int_{0}^{s}ds'\frac{m(s',\Phi,\varphi_0)}{\hbar v_J}}\left(\eta_R(s)-\eta_L(s)\right)\label{eq_appendix: gamma1-end-general},\\ 
    \gamma_2=&\mathcal{N}_2\int_{0}^{L_J} ds e^{\int_{s}^{L_J}ds'\frac{m(s',\Phi,\varphi_0)}{\hbar v_J}}\left(\eta_R(s)+\eta_L(s)\right)\label{eq_appendix: gamma2-end-general}.
\end{align}

Equations (\ref{eq: gamma1-end-zero-B}) and (\ref{eq: gamma2-end-zero-B}) are obtained by applying the condition $\Phi = 0$ to Eqs. (\ref{eq_appendix: gamma1-end-general}) and (\ref{eq_appendix: gamma2-end-general}), where $m(s,\Phi,\varphi_0)=\Delta_0\cos(\varphi_0/2)=m$ is constant along the JJ region $0<s<L_J$.  MZMs appear at the two ends $s=0, L_J$ when $m<0$ (at $\pi<\varphi_0<3\pi$), giving the normalization constant $\mathcal{N}_{1,2}=1/\sqrt{\hbar v_J/|m|}$. 


Equation (\ref{eq: gamma-mobile-B}), the zero energy solution bound to the Josephson vortex when $\Phi\sim \nu \Phi_0$, is found by solving Eq. (\ref{eq_appendix: zero E equation}) under linear approximation of $m(s,\Phi,\varphi_0)$ near the vortex $s=s_n$ where $m(s_n,\Phi,\varphi_0)=0$ \cite{Potter_and_Fu_PRB2013}
\begin{displaymath}
    m(s,\Phi,\varphi_0)= \Delta_0 \cos\left(\frac{\pi \Phi x}{L_J\Phi_0}-\frac{\varphi_0}{2}\right)\approx \frac{\pi\Delta_0 \Phi}{L_J \Phi_0}(-1)^{n}(s-s_n)\, .
\end{displaymath}
The vortex position $s_n$ is given by
\begin{displaymath}
    s_n=\left(n-\frac{1}{2}+\frac{\varphi_0}{2\pi}\right)\frac{L_J}{\Phi/\Phi_0},\,\,\,n=0,\pm 1, \pm 2, ...
\end{displaymath}
The zero energy solution basis state bound to the Josephson vortex is then 
\begin{displaymath}
    \psi_{0,n}=\frac{1}{\sqrt{2}}\begin{pmatrix}
        1\\(-1)^{n}
    \end{pmatrix}\frac{e^{-\frac{(s-s_n)^2}{2\lambda_M^2}}}{\sqrt{2\pi\lambda_M^2}},\,\,\,
    \lambda_M=\sqrt{\frac{\hbar v_J L_J}{\pi \Delta_0 \Phi/\Phi_0}} \, ,
\end{displaymath}
provided that $s_n$ is the inside the JJ region $0<s_n<L_J$.
The localization length $\lambda_M$ depends on $\Phi$ threading the JJ. 
This solution satisfies the boundary condition (\ref{eq_appendix: BC}) in the limit $\lambda_M\ll L_J$, and hence provides the MZMs bound to the vortex 
\begin{equation}\label{eq_appendix: Mobile MZM-general}
    \gamma_n =
    \int ds \frac{e^{-\frac{(s-s_n)^{2}}{2\lambda_M^{2}}}}{\sqrt{2\pi \lambda_{M}^{2}}}\frac{(\eta_{R}(s)+(-1)^{n}\eta_{L}(s))}{\sqrt{2}}\, .
\end{equation}
In Eq. (\ref{eq: gamma-mobile-B}), $n=1$ case is considered. In this case with $n=1$ and $\Phi=\Phi_0$, $s>s_1$ is topological ($m(s,\Phi,\varphi_0)<0$). As a result the zero energy end mode described by Eq. (\ref{eq_appendix: gamma2-end-general}) appears near $s=L_J$, resulting in Eq. (\ref{eq: gamma-end-B})

\section{Numerical simulation method}  
This appendix explains the method for discretizing the Hamiltonian in Eq. (\ref{eq: Effect 1D total H}) on a lattice, which is used to obtain the energy levels and the wave functions in Figs. \ref{fig: level and wave function}, \ref{fig: Boundary effect at zero B}, and \ref{fig: Boundary effect at single flux}. 
The usual discretization of the kinetic part $i\partial_s$ in the Hamiltonian 
results in the spectrum with two Dirac cones. 
The doubling is explained by the no-go theorem by Nielsen and Ninomiya \cite{Nielson_Ninomiya_1, Nielson_Ninomiya_2}
that any Hermitian lattice fermion Hamiltonian with
(i) local hopping (i.e., long-distance hopping decay to zero fast enough so that the Fourier transformation of the Hamiltonian is a smooth function of momentum), (ii) translational invariance, and (iii) chirality (or helicicty) always exhibits an even number of Dirac cones. 
For lattice implementation of single Dirac cone, 
at least one of the three conditions must be broken.

To avoid the Fermion doubling, we adopted the "SLAC method" in Refs.\cite{Drell_FDP_PRD1976, ZiXiang_FDP_ScienceAdvances2018}.
This method avoids the doubling by adopting nonlocal hopping. 


\subsection{Outline of the method}\label{appendix: B.1}
In this method, the chiral fermion Hamiltonian $\displaystyle{H=\int ds \psi^{\dag}(s)\left[-i\hbar v\partial_{s}\right]\psi(s)}$ is implemented on the lattice with site number $N$ and lattice constant $a$ by $H\approx-i\hbar v\sum_{n,n'}\psi_{n}^{\dag}D_{n,n'}\psi_{n'}$,
where the matrix element $D_{n,n'}$, representing the hopping from site $n'$ to site $n$, is given by 
\begin{equation}\label{eq_appendix: Dirac cone-matrix rep}
    D_{n,n'}=
    \begin{cases}
        \displaystyle{\frac{\pi(-1)^{n-n'}}{N a\sin\frac{\pi(n-n')}{N}}} & (n\neq n')\\  
        &\\
        0 & (n=n)   
    \end{cases}.
\end{equation} 
The resulting spectrum is exactly linear with a single Dirac point at $k=0$. 

To implement the antiperiodic boundary condition (BC) induced by Berry phase $\pi$ in a round trip, we adopted an even number of lattice site $N$.
An odd value of $N$ cannot realize the antiperiodic BC, since $D_{n,n'}$ with odd $N$ possesses a zero eigenvalue ($k=0$ state) which happens with the periodic BC (
i.e., since the determinant of the matrix $D_{n,n'}$ is always zero when $N$ is odd, due to its antisymmetric property $D_{n,n'}=-D_{n',n}$).
The antiperiodic BC, which possesses nonzero eigenvalues only, is realized only when $N$ is even. 

The appropriate value of the lattice constant $a$
was chosen by 
monitoring the convergence of the energy levels with decreasing $a$.
The value $a=6.58$nm is selected, where proper convergence of energy level $E$ within $E\in [-\Delta_0,\Delta_0]$ was achieved in all value of $\varphi_0$. 

\subsection{Derivation of the Eq. \eqref{eq_appendix: Dirac cone-matrix rep}}\label{appendix: B.2}
We derive Eq. \eqref{eq_appendix: Dirac cone-matrix rep}, following Refs. \cite{Drell_FDP_PRD1976,ZiXiang_FDP_ScienceAdvances2018}. 
For the derivation, we apply the first Brillouin zone (FBZ) $- \pi/a\leq k<\pi/a$, with the momentum cut-off $\pi/a$,
to 
the Fourier space expression of the Hamiltonian \cite{Drell_FDP_PRD1976}, 
\begin{displaymath}
    H\approx \sum_{k\in \text{FBZ}}\hbar v k\psi_{k}^{\dag}\psi_{k}=-i\hbar v\sum_{n',n}\sum_{k\in FBZ}\psi_{n}^{\dag}\left[\frac{ik}{N}e^{ika(n-n')}\right]\psi_{n'},
\end{displaymath}
where $\psi_{k}=\frac{1}{\sqrt{N}}\sum_{n''}\psi_{n''}e^{-ikn''a}$. 
As a result, we find
\begin{equation}\label{eq_appendix: Dirac cone-Fourier def}
    D_{n,n'}=\frac{1}{N}\sum_{k\in \text{FBZ}}ike^{ika(n-n')}\,.
\end{equation}

To perform the sum over $k$ in Eq. \eqref{eq_appendix: Dirac cone-Fourier def}, the values of momentum $k$ must be specified from the BC of the system.
The BC is imposed by $\psi_{n+N}=e^{i\zeta}\psi_n$ 
with $\zeta = \pi$ for the anti-periodic BC (in the presence of the Berry phase $\pi$) and $\zeta = 0$ for the periodic BC (in the absence).
The momentum values satisfying the BC are given by $k=\frac{2\pi \nu  +\zeta}{Na}$, with integer values of $\nu$, 
$\nu=\nu_{\text{min}},\nu_{\text{min}}+1, ... ,\nu_{\text{min}}+N-1$ for the FBZ.
Here, $\nu_{\text{min}}$ satisfies $k_{\text{min}}=\frac{2\pi \nu_{\text{min}}+\zeta}{Na}$, and $k_{\text{min}}$ is the minimum value of the momentum in the FBZ satisfying the BC. 

When $n=n'$, the sum over $k$ in Eq. \eqref{eq_appendix: Dirac cone-Fourier def} becomes
\begin{align}
    D_{n,n}=&\sum_{k\in \text{FBZ}}\frac{ik}{N}=i\sum_{\nu=\nu_{\text{min}}}^{\nu_{\text{min}+N-1}}\left(\frac{2\pi\nu+\zeta}{N^2 a}\right)\nonumber\\ 
    =& i\left(k_{\text{min}}+\frac{\pi(N-1)}{Na}\right).\nonumber
\end{align}
For $i D_{n,n'}$ to be a Hermitian matrix, we have $D_{n,n'}=-D_{n',n}$, namely $D_{n,n}=0$. 
Hence
\begin{equation}\label{eq_appendix: Dirac cone_zero diagonal cond}
    k_{\text{min}} + \frac{\pi}{a}\left(1-\frac{1}{N}\right)=0.
\end{equation}
Comparing Eq. \eqref{eq_appendix: Dirac cone_zero diagonal cond} with $k_{\text{min}} a=(2\pi\nu_{\text{min}}+\zeta)/N$,
we have
$N=-2\nu_{\text{min}}-1$ (which is odd) for the periodic BC $\zeta=0$, and $N=-2\nu_{\text{min}}$ (which is even) for the antiperiodic $\zeta=\pi$. 
This shows that the even-oddness of the total number $N$ of lattice sites is constrained by the BC.

For $n\neq n'$, $D_{n,n}$ is given by
\begin{align}
    D_{n,n}=&\frac{1}{Na}\partial_n\sum_{k\in \text{FBZ}}e^{ika(n-n')}\nonumber\\ 
    =&\partial_{n}\left[\frac{e^{ik_{\text{min}}a(n-n')}}{Na}\sum_{\nu=\nu_{\text{min}}}^{\nu_{\text{min}}+N-1}e^{\frac{2\pi i (\nu-\nu_{\text{min}})}{N}(n-n')}\right]\nonumber\\ 
    =&e^{i\left(k_{\text{min}} a + \pi\left(1-\frac{1}{N}\right)\right)(n-n')} \partial_{n} \left( \frac{\sin(\pi(n-n'))}{Na\sin\left(\frac{\pi(n-n')}{N}\right)}\right) \nonumber\\
    &+ i\left(k_{\text{min}} a + \frac{\pi}{a}\left(1-\frac{1}{N}\right)\right) \frac{\sin(\pi(n-n'))}{Na\sin\left(\frac{\pi(n-n')}{N}\right)}\nonumber
\end{align}
Using Eq. \eqref{eq_appendix: Dirac cone_zero diagonal cond}, Eq. \eqref{eq_appendix: Dirac cone-matrix rep} is obtained.

\end{document}